\documentclass[10pt, twocolumn]{article}

\usepackage{psfig}
\usepackage{latexsym}
\usepackage{amssymb}
\usepackage{epsfig,amsmath,graphics}

\newcommand{\OO}{\mathcal{O}}
\newcommand{\LL}{\mathcal{L}}

\newcommand{\bino}{\tilde{B}}
\newcommand{\wino}{\tilde{W}}
\newcommand{\gluino}{\tilde{g}}
\newcommand{\higgsino}{\tilde{H}}
\newcommand{\Top}{\text{top}}

\begin{document}
\twocolumn[
\begin{center}
{\large\bf One Loop Predictions of the Finely Tuned SSM}\\
\vskip 0.3cm
{\normalsize
Asimina Arvanitaki, Chad Davis, Peter W. Graham, and Jay G. Wacker\\
\vskip 0.2cm
Institute for Theoretical Physics\\
Department of Physics\\
Stanford University\\
Stanford, CA 94305 USA\\
\vskip .1in}
\end{center}

\vskip .5cm

\begin{abstract}
We study the finely tuned SSM, recently proposed by Arkani-Hamed and
Dimopoulos, at the one loop level. The runnings of the four gaugino
Yukawa couplings, the $\mu$ term, the gaugino masses, and the Higgs
quartic coupling are computed. The Higgs mass is found to be 130 --
170 GeV for $M_s
>  10^6$ GeV. Measuring the Yukawa coupling constants at the
10\% level can begin to constrain the SUSY breaking scale. Measuring
the relationships between the couplings will provide a striking
signal for this model.
\end{abstract}
\vskip 1.0cm
]

\section{Introduction}

Recently there has been interest in studying a version of the
Supersymmetric Standard Model (SSM) where naturalness is no longer a
guiding principle \cite{Arkani-Hamed:2004fb}.  This comes at a time
of several growing problems associated with the standard
implementation of naturalness \cite{Barbieri:2000gf}.  The most
pressing naturalness issue is the cosmological constant, the
experimental value of which appears to be fine-tuned to one part in
$10^{120}$ and completely dwarfs the standard hierarchy problem.
While it is conceivable that these two separate fine tunings are
divorced, they could also be linked with weak anthropicism
\cite{Weinberg:dv}. There are other problems with the SSM directly
related to particle physics issues, such as the non-discovery of
superpartners at LEP or Fermilab, the lack of FCNCs, the
non-discovery of the Higgs, and the non-discovery of proton decay.
All of these increase the fine-tuning required in the SSM. Every one
of these phenomenological problems is ameliorated by decoupling the
scalars \cite{Arkani-Hamed:2004fb,Wells:2003tf}.

The two major successes of the SSM \cite{Dimopoulos:1981zb} are
gauge coupling unification \cite{Dimopoulos:1981yj} and a viable
dark matter candidate. However, removing the scalars of the SSM does
not significantly alter either of these predictions. If one is
willing to ignore the original motivation for the SSM and decouple
all but the one scalar Higgs doublet required for electroweak
symmetry breaking, then one immediately has a phenomenologically
viable model without the usual concerns of the SSM. The existence of
light gauginos and Higgsinos is inferred indirectly through gauge
coupling unification and evidence for dark matter, which point to
these states having mass in the 100 GeV to 3 TeV range.

There is a universal form of the low energy effective action for the
finely tuned SSM that preserves gauge coupling unification and dark
matter and has five relevant interactions -- four Yukawa couplings
from the gauginos and the Higgs quartic coupling.  These are
predicted by high energy supersymmetry from four parameters: the
Standard Model gauge couplings $g_1$ and $g_2$, $\tan \beta$, and
the scale of the scalar masses, $M_s$. At the LHC or NLC it may be
possible to measure five new couplings and explain them from only
two new parameters.

In this note, we calculate the one loop beta functions of these five
couplings, as well as those of the $\mu$ term and the gaugino
masses. We then run these couplings from their SUSY values at $M_s$
down to the top mass $m_t$
\cite{Olechowski:1988ol,Chankowski:1990ch,Casas:1994ca,Haber:1996fp}.
We do not compute threshold corrections because they are subdominant
to the large logarithms. We define two different effective $\tan
\beta$ that are related to the gaugino-Higgsino Yukawa coupling.  By
RG evolving these to a higher scale it is possible to determine the
scale of SUSY breaking.

\section{One loop beta functions}

The tree level Lagrangian contains the terms 
\begin{eqnarray}
\nonumber
\LL &\supset&  \bino (\kappa'_1 h^\dagger \higgsino_1 + \kappa_2' h
\higgsino_2)\\
&& +\wino^a( \kappa_1 h^\dagger \tau^a \higgsino_1 + \kappa_2
\higgsino_2 \tau^a h) - \lambda |h|^4\\
\nonumber&& - \mu \higgsino_1 \higgsino_2 - \frac{1}{2}(M_1
\bino\bino + M_2 \wino\wino + M_3 \gluino\gluino).
\end{eqnarray}
At the SUSY breaking scale the following relations are satisfied: 
\begin{eqnarray}
\nonumber &&\kappa'_1= \sqrt{\frac{3}{10}} g_1 \sin \beta
\hspace{0.3in} \kappa'_2= \sqrt{\frac{3}{10}} g_1 \cos\beta
  \\
\nonumber &&\kappa_1= \sqrt{2} g_2 \sin \beta \hspace{0.3in}
\kappa_2= \sqrt{2} g_2 \cos \beta\\
&&\lambda = \frac{\frac{3}{5}g_1^2+g_2^2}{8}\cos ^2 2 \beta.
\end{eqnarray}
However, these couplings run in a non-supersymmetric fashion from
the SUSY breaking scale down to low energies.

All of the following results are given with SU(5) normalization of
the hypercharge.  The beta function for the Higgs quartic coupling
is 
\begin{eqnarray}
\nonumber 16\pi^2 \beta_\lambda &=& + 24 \lambda^2 - 6 y_{\Top}^4 +
12 \lambda y_{\Top}^2\\
\nonumber&& + \frac{27}{200} g_1^4 + \frac{9}{20} g_1^2
g_2^2\\
\nonumber && + \frac{9}{8} g_2^4 - \frac{9}{5} \lambda g_1^2 - 9
\lambda g_2^2\\
\nonumber&&  - \frac{5}{8} (\kappa^4_1 + \kappa^4_2) - \frac{1}{4}
\kappa^2_1 \kappa^2_2\\
\nonumber&& - 2 (\kappa'^2_1 +\kappa'^2_2)^2 - (\kappa_1 \kappa'_1 +
\kappa_2 \kappa'_2)^2
\\
&& + 3\lambda (\kappa_1^2+ \kappa_2^2) + 4 \lambda(\kappa_1'{}^2 +
\kappa_2'{}^2).
\end{eqnarray}
The beta function for the top Yukawa coupling is
\begin{eqnarray}
\nonumber
16\pi^2\beta_{y_{\Top}} &=& \frac{9}{2} y_{\Top}^3
  -y_{\Top}(\frac{17}{20} g_1^2 + \frac{9}{4} g_2^2 + 8
g_3^2)\\
\nonumber && +\frac{3}{4} y_{\Top} (\kappa_1^2 +\kappa_2^2)\\
&& + y_{\Top} (\kappa_1'{}^2 + \kappa_2'{}^2).
\end{eqnarray}
As a check, when all $\kappa$'s are set to zero, these $\beta$
functions reproduce those of the Standard Model \cite{Arason:1992}.
The beta function for the bino Yukawa coupling is 
\begin{eqnarray}
\nonumber
16\pi^2\beta_{\kappa'_1} &=& 3\kappa'_1 y_{\Top}^2 -
\kappa'_1(\frac{9}{20} g_1^2 + \frac{9}{4}g_2^2)\\
\nonumber
&& + \frac{5}{2} \kappa'_1{}^3
  + 4 \kappa'_1 \kappa'_2{}^2
  + \frac{9}{8}\kappa'_1 \kappa_1^2\\
  &&
  + \frac{3}{4} \kappa'_1 \kappa_2^2 + \frac{3}{2} \kappa_1 \kappa_2
\kappa'_2
\end{eqnarray}
and similarly for $\kappa_2'$ after changing
$\kappa_1\leftrightarrow \kappa_2$ and
$\kappa'_1\leftrightarrow\kappa'_2$.

The beta function for the wino Yukawa coupling is
\begin{eqnarray}
\nonumber
16\pi^2 \beta_{\kappa_1} &=&
3 y_{\Top}^2 \kappa_1 - \kappa_1(\frac{9}{20}g_1^2 + \frac{33}{4} g_2^2)\\
\nonumber
&&+ \frac{11}{8} \kappa_1^3 +\frac{3}{2} \kappa_1 \kappa'_1{}^2 +
\frac{1}{2} \kappa_1 \kappa_2^2\\
&& +\kappa_1 \kappa'_2{}^2 + 2 \kappa'_1 \kappa_2 \kappa'_2
\end{eqnarray}
and similarly for $\kappa_2$ after changing $\kappa_1\leftrightarrow
\kappa_2$ and $\kappa'_1\leftrightarrow\kappa'_2$.

The beta function for the $\mu$ term is 
\begin{eqnarray}
\nonumber 16\pi^2 \beta_{\mu} &=&
- \mu(\frac{9}{10}g_1^2 + \frac{9}{2} g_2^2)\\
\nonumber &&+ \frac{3}{2} \kappa_1\kappa_2 M_2 +2 \kappa'_1
\kappa'_2 M_1\\
&&+ \frac{3}{8}\mu (\kappa_1^2 + \kappa_2^2) + \frac{1}{2} \mu
(\kappa'_1{}^2 + \kappa'_2{}^2).
\end{eqnarray}

The beta functions for the gaugino masses are 
\begin{eqnarray}
16\pi^2 \beta_{M_1} &=&
8 \mu\kappa'_1\kappa'_2 + 2 M_1 (\kappa'_1{}^2 + \kappa'_2{}^2)\\
16\pi^2 \beta_{M_2}&=& -12 g_2^2 M_2 + 2 \mu\kappa_1\kappa_2\\
\nonumber&& + \frac{1}{2}M_2(\kappa_1^2 + \kappa_2^2)\\
16\pi^2 \beta_{M_3} &=& -18 g_3^2 M_3.
\end{eqnarray}

In the following sections we run the Yukawa couplings and the mass
terms from the SUSY breaking scale down to the low scale. We examine
the behavior of the various parameters at the low scale as a
function of $M_s$.

\begin{figure}
\begin{center}
\epsfig{file=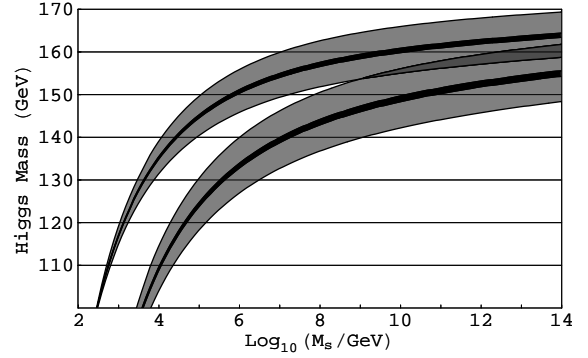, width=3.0in} \caption{ \label{Fig: Higgs
Mass} The Higgs mass as a function of the SUSY breaking scale
$\log_{10}(M_s/\text{GeV})$.  The upper bands are for
$\tan\beta(M_s) = 50$ and the lower ones are $\tan\beta(M_s) = 1$.
The width of each grey band is the experimental uncertainty, mainly
due to $m_t$. The width of each black band is the uncertainty when
expected improvements from a future linear collider are taken into
account.}
\end{center}
\end{figure}

\section{Higgs Mass}

The Higgs quartic coupling at $M_s$ depends only on  $\cos 2\beta$
and $M_s$ and can easily be run down with the beta functions of the
previous section. We find that the Higgs is heavier than in the
usual SSM with low-scale SUSY breaking
\cite{Haber:1996fp,Okada:1990vk}. The dimensionful A-terms and $\mu$
term are around the weak/dark matter scale and are small in
comparison to the SUSY breaking scale. They give finite threshold
effects to the Higgs quartic coupling that are $\OO(A^2/M_s^2)$ and
can be neglected in this model. We have used a top mass of $178.0
\pm 4.3$ GeV \cite{Group:2004rc}. The $\overline {MS}$ top Yukawa
coupling was set to $y_t = 0.99 \pm 0.02$ by the relation
\cite{Haber:1996fp,Arason:1992}
\begin{eqnarray}
m_t= y_t v (1 +\frac{16}{ 3} \frac{g_3^2}{16\pi^2} - 2
\frac{y_t^2}{16\pi^2}).
\end{eqnarray}

For a SUSY breaking scale of $10^9$ GeV, we find that the Higgs mass
varies from 140 to 165 GeV as $\cos 2\beta$ goes from 0 to 1 at the
high scale.  The Higgs mass as a function of $M_s$ is shown in Fig.
1 for $\tan\beta = 1$ and $\tan\beta = 50$. For values of
$\tan\beta$ between 1 and 50, the Higgs mass is between the bounds
shown.  The Higgs quartic coupling is insensitive to $\tan\beta$ for
large $\tan\beta$.

Experimental uncertainties in $y_t$ and $g_3$ lead to an uncertainty
in the prediction of the Higgs mass as shown by the wide bands in
Fig. 1. The error in the top mass dominates while the uncertainty
due to $g_3$ is approximately one tenth as large. As a test of the
theoretical uncertainty, each $\mathbf{5} \oplus \mathbf{\bar{5}}$
fermion added in at the TeV scale increases the Higgs mass by 0.2\%
for $M_s = 10^9$ GeV.

A future linear collider may be able to measure the Higgs mass to a
precision of 100 MeV, the top mass to 200 MeV, and $\alpha_s$ to 1\%
\cite{Abe:2001ab}.  The narrow bands in Fig. 1 show the uncertainty
in the Higgs mass prediction using these more precise measurements
and assuming the current central value. The small error on the Higgs
mass measurement could allow the most precise determination of the
SUSY breaking scale within the context of this model.  If
$\tan\beta$ is measured to $50\%$, $M_s$ will be known to within an
order of magnitude. Although the bands in Fig. 1 asymptote at high
scales making $M_s$ difficult to determine from the Higgs mass, we
do not expect $M_s$ to be greater than $10^{13}$ GeV
\cite{Arkani-Hamed:2004fb}.

\section{Yukawa Couplings and Mass Terms}

\begin{figure}
\begin{center}
\epsfig{file=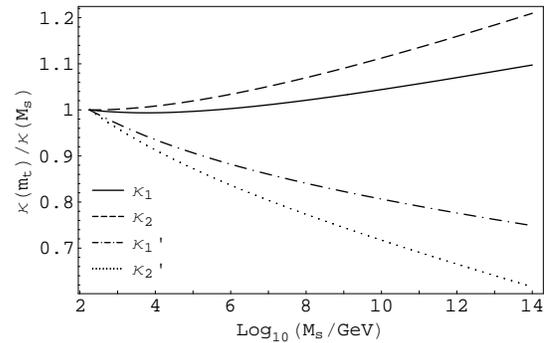, width=3.0in} \caption{ \label{Fig:
Kappa1} The ratio $\kappa(m_t)$/$\kappa(M_s)$ as a function of $M_s$
for fixed $\tan\beta(M_s)=5$. }
\end{center}
\end{figure}

\begin{figure}
\begin{center}
\epsfig{file=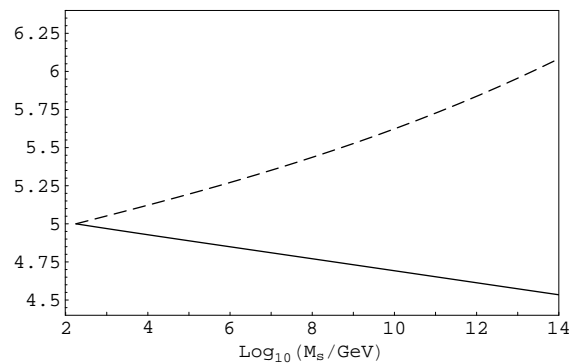, width=3.0in} \caption{ \label{Fig:
TanBeta} The solid line shows $\tan\beta_{\text{low}}(m_t)$ as a
function of $M_s$.  The dashed line is for
$\tan\beta'_{\text{low}}(m_t)$. Here $\tan\beta(M_s) = 5$. }
\end{center}
\end{figure}

\begin{figure}
\begin{center}
\epsfig{file=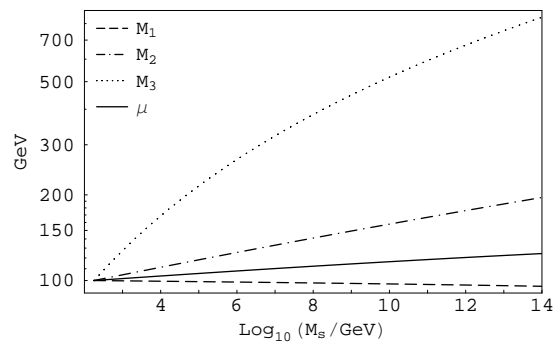, width=3.0in} \caption{
\label{Fig:gaugino} The gaugino masses and $\mu$ evaluated at $m_t$
as a function of $M_s$ for fixed $\tan\beta(M_s) = 5$.}
\end{center}
\end{figure}

The gaugino couplings are set at $M_s$ by Eq. (2) and RG evolved to
$m_t$.  There are two separate low energy definitions of $\tan
\beta$, 
\begin{eqnarray}
\tan\beta_{\text{low}}(m) = \frac{\kappa_1(m)}{\kappa_2(m)}
\hspace{0.3in} \tan\beta'_{\text{low}}(m) =
\frac{\kappa'_1(m)}{\kappa'_2(m)},
\end{eqnarray}
that run from equal values at the SUSY breaking scale.  Running up
from the weak scale to the point where they unify provides a clear
determination of the SUSY breaking scale (Fig.3).  If the couplings
could be measured to $10\%$ at a future LC \cite{Danielson:1996tf}
this would determine $M_s$ to within a few orders of magnitude. Note
that there are fixed points in the evolution of some of the
$\tan\beta$'s at $\tan\beta = 0, 1, \infty$. However, the gaugino
couplings do change as $M_s$ is changed and therefore can provide a
useful measure of $M_s$ even when $\tan\beta_{\text{low}}$ does not
change significantly with $M_s$.

The Yukawa couplings run significantly from their supersymmetric
values (Fig.2). We find that, for $\tan\beta \gtrsim 5$, the ratios
$\kappa(m_t)/\kappa(M_s)$ are relatively unaffected by changes in
$\tan\beta$. The four Yukawa couplings and the Higgs quartic are
five independently measurable parameters that are determined by the
scale of SUSY breaking and $\tan\beta$. Thus, this model predicts
that these five couplings will satisfy three relations at the low
scale.

Finally, using the calculated $\beta$ functions, the running values
of $\mu$ and the gaugino masses can be found.  As a simple example,
we set all four masses equal to 100 GeV at $M_s$ and then run them
down to the low scale. As shown in Fig.4, the gluino mass increases
greatly at the low scale (to $\sim 400$ GeV for $M_s = 10^9$ GeV).
$M_2$ and $\mu$ increase modestly while $M_1$ decreases slightly.
The running of the gluino mass depends only on $g_3$ and $M_3$, so
the ratio $M_3(m_t)/M_3(M_s)$ is independent of the specific values
chosen for $M_3$ and $\tan\beta$. Although the runnings of $\mu$,
$M_1$, and $M_2$ are more complicated, they are relatively
insensitive to changes of $\tan\beta$. We expect $\mu$ and the
gaugino masses to be of the same order as the weak scale
\cite{Arkani-Hamed:2004fb}.

\section{Conclusion}

We have computed the one loop leading log running for the finely
tuned SSM where the scalars are much heavier than the weak scale. We
find that the Higgs mass is in the 140 to 165 GeV range at $M_s =
10^9$ GeV, depending on $\tan \beta$. The Higgs mass should be
calculated at the two loop level including one loop threshold
effects for a more exact prediction. The gaugino Yukawa couplings
were found to run significantly, and, if measured to ten percent
accuracy at an NLC, could determine the scale of SUSY breaking to
within a few orders of magnitude.  A measurement of the Higgs mass
and $\tan\beta$ could provide an even better estimate of $M_s$, but
does not verify the model. More work is needed to determine how
effectively the LHC and NLC will be able to extract the gaugino
Yukawa couplings, but a measurement of the relationships between
these couplings would provide a phenomenal signal of high scale
supersymmetry.

\section*{Acknowledgements}
We would like to thank N. Arkani-Hamed, S. Dimopoulos, S. Martin, A.
Pierce, and S. Thomas for useful discussions.  Special thanks to
G.F. Giudice and A. Romanino for pointing out errors in our
calculation \cite{Giudice:2004gr}.  Our results are now in
agreement.  C.D. is the Mellam Family Graduate Fellow.  P.W.G. is
supported by the National Defense Science and Engineering Graduate
Fellowship. J.G.W. is supported by NSF grant PHY-9870115 and the
Stanford Institute for Theoretical Physics.

\providecommand{\href}[2]{#2}\begingroup\raggedright

\endgroup


\begin{thebibliography}{1}

\bibitem{Arkani-Hamed:2004fb}
N.~Arkani-Hamed and S.~Dimopoulos,
arXiv:hep-th/0405159.

\bibitem{Barbieri:2000gf}
R.~Barbieri and A.~Strumia,
arXiv:hep-ph/0007265.



\bibitem{Weinberg:dv}
S.~Weinberg,
Phys.\ Rev.\ Lett.\  {\bf 59}, 2607 (1987);
T.~Banks, Nucl. Phys. B \textbf{249}, 332 (1985);
A.~D.~Linde,
Mod.\ Phys.\ Lett.\ A {\bf 1}, 81 (1986);
S.~Kachru, R.~Kallosh, A.~Linde and S.~P.~Trivedi,
Phys.\ Rev.\ D {\bf 68}, 046005 (2003) [arXiv:hep-th/0301240];
L.~Susskind,
arXiv:hep-th/0302219;
M.~R.~Douglas, JHEP \textbf{0305}, 046 (2003)[arXiv:hep-ph/0401004];
F.~Denef and M.~R.~Douglas, arXiv:hep-th/0404116.

\bibitem{Wells:2003tf}
J.~D.~Wells,
arXiv:hep-ph/0306127.

\bibitem{Dimopoulos:1981zb}
S.~Dimopoulos and H.~Georgi,
Nucl.\ Phys.\ B {\bf 193}, 150 (1981).

\bibitem{Dimopoulos:1981yj}
S.~Dimopoulos, S.~Raby and F.~Wilczek,
Phys.\ Rev.\ D {\bf 24}, 1681 (1981).


\bibitem{Olechowski:1988ol}
M.~Olechowski and S.~Pokorski,
Phys. Lett. B {\bf 214}, 393 (1988).

\bibitem{Chankowski:1990ch}
P.~H.~Chankowski,
Phys.\ Rev.\ D {\bf 41}, 2877 (1990).

\bibitem{Casas:1994ca}
J.~A.~Casas,
arXiv:hep-ph/9407389.

\bibitem{Haber:1996fp}
H.~E.~Haber, R.~Hempfling and A.~H.~Hoang,
Z.\ Phys.\ C {\bf 75}, 539 (1997) arXiv:hep-ph/9609331.

\bibitem{Arason:1992}
H.~Arason, D.~J.~Castano, B.~Kesthelyi, S.~Mikaelian, E.~J.~Piard,
P.~Ramond, and B.~D.~Wright,
Phys.\ Rev.\ D {\bf 46}, 3945 (1992).

\bibitem{Okada:1990vk}
Y.~Okada, M.~Yamaguchi and T.~Yanagida,
Prog.\ Theor.\ Phys.\  {\bf 85}, 1 (1991);
H.~E.~Haber and R.~Hempfling,
Phys.\ Rev.\ Lett.\  {\bf 66}, 1815 (1991);
J.~R.~Ellis, G.~Ridolfi and F.~Zwirner,
Phys.\ Lett.\ B {\bf 257}, 83 (1991).

\bibitem{Group:2004rc}
The CDF Collaboration, the D0 Collaboration and the Tevatron
Electroweak Working Group,
arXiv:hep-ex/0404010.

\bibitem{Abe:2001ab}
T.~Abe et. al.
arXiv:hep-ex/0106055, arXiv:hep-ex/0106056, arXiv:hep-ex/0106057,
arXiv:hep-ex/0106058.

\bibitem{Danielson:1996tf}
M.~N.~Danielson {\it et al.},
eConf {\bf C960625}, SUP117 (1996).

\bibitem{Giudice:2004gr}
G.~F.~Giudice and A.~Romanino,
arXiv:hep-ph/0406088.

\end{thebibliography}
\end{document}